\documentclass[twocolumn,american,prb,showpacs,superscriptaddress]{revtex4-1}
\usepackage{amsmath}
\usepackage{graphicx}
\usepackage{amssymb}
\usepackage{times}
\usepackage{color}
\usepackage[flushleft]{threeparttable}
\definecolor{lr}{rgb}{1.0,0.3,0.3}
\definecolor{dg}{rgb}{0.0,0.5,0.0}

\makeatletter

\@ifundefined{textcolor}{}
{%
 \definecolor{BLACK}{gray}{0}
 \definecolor{WHITE}{gray}{1}
 \definecolor{RED}{rgb}{1,0,0}
 \definecolor{GREEN}{rgb}{0,1,0}
 \definecolor{BLUE}{rgb}{0,0,1}
 \definecolor{CYAN}{cmyk}{1,0,0,0}
 \definecolor{MAGENTA}{cmyk}{0,1,0,0}
 \definecolor{YELLOW}{cmyk}{0,0,1,0}
 }

\makeatother

\begin{document}

\title{An \emph{ab initio} study on split silicon-vacancy defect in diamond:
  electronic structure and related properties}

\author{Adam Gali} 
\email{agali@eik.bme.hu}
\affiliation{Wigner Research Centre for Physics, Hungarian Academy of Sciences,
  PO Box 49, H-1525, Budapest, Hungary}
\affiliation{Department of Atomic Physics, Budapest University of
  Technology and Economics, Budafoki \'ut 8., H-1111 Budapest,
  Hungary}


\author{Jeronimo R. Maze}
\email{jeromaze@gmail.com}
\affiliation{Faculty of Physics, Pontificia Universidad Cat\'olica de Chile, Santiago 7820436, Chile}

\pacs{61.72.J-, 61.82.Fk, 71.15.Mb, 76.30.-v}

\begin{abstract}
The split silicon-vacancy defect (SiV) in diamond is an electrically and
optically active color center. Recently, it has been shown that this
color center is bright and can be detected at the single defect level. In addition, the SiV
defect shows a non-zero electronic spin ground state that potentially makes this
defect an alternative candidate for quantum optics and metrology applications beside
the well-known nitrogen-vacancy color center in diamond. However, the
electronic structure of the defect, the nature of optical excitations
and other related properties are not well-understood. Here we present
advanced \emph{ab initio} study on SiV defect in
diamond. We determine the formation energies, charge transition levels and the nature of excitations of the defect. Our study unravel the origin of the dark or shelving state for the negatively charged SiV defect associated with the 1.68-eV photoluminescence center.
\end{abstract}
\maketitle

\section{Introduction}
\label{Introduction}

The 1.68-eV photoluminescence (PL) center was reported many decades ago in
diamond\cite{Vavilov80} and where later it was assumed \cite{Zaitsev81}
that silicon impurities were involved in this center. This was
confirmed by PL measurements at cryogenic temperature on Si-doped
chemical vapor deposition (CVD) polycrystalline diamond samples where
the fine structure of the 1.68-eV PL center could be
detected.\cite{Clark95} A 12-line fine structure is observed close to
1.68~eV, and this can be divided into three similar groups each
containing four components. The relative strengths of the optical
absorption for the three groups of lines are found to be the same as
the ratio of the abundancies of the natural isotopes of silicon,
$^{28}$Si, $^{29}$Si, and $^{30}$Si\cite{Clark95}. The 4-line fine structure for an
individual Si-isotope is assigned to doublet levels both in the ground
and excited states which split by about 48 and 242~GHz,
respectively.\cite{Clark95} It has been assumed that this small
splitting might be explained by dynamic Jahn-Teller
effect\cite{Clark95} and/or by spin-orbit effect.\cite{Goss96}

Recently, single photon-emission from 1.68-eV PL center has been
demonstrated \cite{Wang06, Neu11, Neu11PRB, Neu13}. Its zero-phonon-line (ZPL)
with 5~nm width even at room temperature and the near infrared
emission makes this PL center very attractive candidate for quantum
optics\cite{Wang06,Neu11,Neu11PRB} and biomarker\cite{Barnard09} applications.

\begin{figure}
\includegraphics[width=0.95\columnwidth]{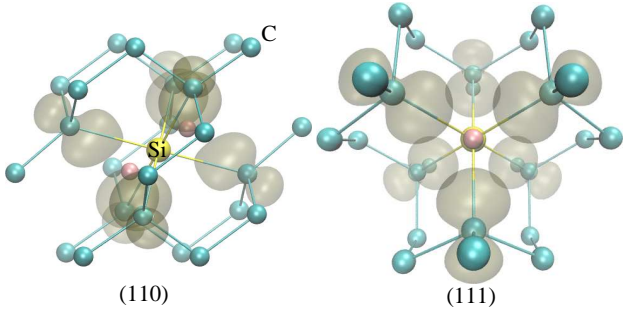} 
\caption{\label{fig:struct}(Color online) The split-vacancy structure
  of the SiV defect and the calculated spin density in neutral charge
  state. The isosurface of spin density is 0.05. The defect has
  $D_{3d}$ symmetry with the symmetry axis of [111] direction. The lattice sites of the missing carbon atoms are depicted by the smallest (pink) balls.
}
\end{figure}

Spin-polarized local density approximation\cite{C-A80, Perdew81} (LDA)
within density functional theory (DFT) calculations in a very small
nanodiamond model (with $\sim$70 carbon atoms) concluded that the
negatively charged split-vacancy form (see Fig.~\ref{fig:struct}) of
SiV defect (SiV$^{-}$) is responsible for the 1.68-eV PL
center.\cite{Goss96} They exclude the neutral SiV defect (SiV$^{0}$)
as a good candidate as its ground state is orbitally
singlet.\cite{Goss96} This model was later disputed based on a
semiempirical restricted open-shell Hartree-Fock cyclic cluster
model calculation where they claimed that tunneling of Si-atom along
the symmetry axis may occur for SiV$^{0}$ defect that can explain the
doublet line in the ground state.\cite{Moliver03}    

The fingerprint of SiV$^{0}$ was found by electron paramagnetic
resonance (EPR) studies.\cite{Iakoubovskii00, Iakoubovskii02, Edmonds08} The
KUL1 center with $S=1$ high spin ground state and D$_{3d}$
symmetry\cite{Edmonds08} was recently associated with SiV$^0$ defect
where 216-atom LDA supercell calculations produces relatively good
agreement with the measured $^{13}$C and $^{28}$Si hyperfine couplings
(see Tab.~\ref{tab:hf}). Very recently, thorough EPR and PL studies
have been carried out to correlate the KUL1 EPR center with an 1.31-eV
PL center, and its relation to the 1.68-eV PL center.\cite{DHJ11} The
final conclusion was that SiV$^0$ has a ZPL at 1.31-eV whereas
SiV$^{-}$ has a ZPL at 1.68-eV. Photo-conductivity
measurements\cite{Allers95} and photo-ionization
measurements\cite{DHJ11} indicate that the adiabatic (thermal) charge
transition level of $(-|0)$ level of SiV defect is at $\sim
E_\text{VBM}$+1.5~eV, where VBM is the valence band edge.

While the recent measurements\cite{DHJ11} are very plausible still no
\emph{ab initio} theory was able to conclusively support the
assignment of 1.68-eV center with the negatively charged SiV
defect. LDA or any semilocal generalized gradient approximation (GGA)
functionals suffer from the band gap error\cite{Perdew81, Gali11}
which inhibits to directly compare the calculated and experimental ZPL
energies. Recent advances in DFT functionals made possible to
accurately calculate ZPL energies and charge transition level of defects in
diamond and other semiconductors.\cite{Gali09, Deak10,Deak13} In this paper,
we apply this theory to study the charge transition levels and the ZPL
energies of SiV defect in diamond. These calculations yield the
position and nature of defect states in host diamond and are able to reveal the 
nature of the shelving state in 1.68-eV PL center. 

Our paper is organized as follows. In Section~\ref{sec:method} we
describe briefly the \emph{ab initio} method that we applied to study
the electronic structure and excitations of the SiV defect. In
Section~\ref{sec:results} we describe the structure and the basic
defect level scheme of SiV defect by group theory. Here, we combine
the results from \emph{ab initio} calculations with group theory
considerations in order to identify the order of important defect
states and the charge state relevant for the most important 1.68-eV PL
center. We discuss the results then we conclude and
summarize the results in Section~\ref{sec:summary}.

\section{Method}
\label{sec:method} 

The present calculations have been carried out in the framework of the
generalized Kohn-Sham theory,\cite{Fuchs07} by using the screened
hybrid functional HSE06 of Heyd, Ernzerhof and Scuseria with the
original parameters (0.2~\AA $^{-1}$ for screening and 25\%
mixing).\cite{Heyd03, Krukau06} HSE06 in diamond happens to be nearly
free of the electron self-interaction error, and is capable of
providing defect-levels and defect-related electronic transitions
within $\sim$0.1~eV to experiment.\cite{Gali09, Deak10,Deak13}

We have used the Vienna Ab Initio Simulation Package VASP~5.3.2 with
the projector augmented wave\cite{PAW} (PAW) method (applying
projectors originally supplied to the 5.2 version).\cite{VASP} To
avoid size effects as much as possible, a 512-atom supercell was used
in the $\Gamma$-approximation for defect studies. Parameters for the
supercell calculations were established first by using the GGA
exchange of Perdew, Burke and Ernzerhof (PBE)\cite{PBE} in bulk
calculations on the primitive cell with a 8$\times$8$\times$8
Monkhorst-Pack (MP) set for Brillouin-zone sampling.\cite{MP76}
(Increasing the MP set to 12$\times$12$\times$12 has changed the total
energy by $<$0.002~eV.) Constant volume relaxations using a cutoff of
370(740)~eV in the plane-wave expansion for the wave function (charge
density) resulted in an equilibrium lattice parameter of $a
_\text{PBE} $= 3.570~\AA . Increasing the cutoff to 420 (840) eV has
changed the lattice constant by only 0.003~\AA. Therefore, considering
the demands of the supercell calculations, the lower cut-off was
selected. The HSE06 calculation with the 8$\times$8$\times$8 MP set
and 370(740)~eV cutoff resulted in the lattice constant
$a_\text{HSE}$=3.545~\AA , indirect band gap of $E_\text{g}$=5.34~eV,
in good agreement with the experimental values of $a$=3.567~\AA\ and
$E_\text{g}$=5.48~eV (see, e.g., Ref.~\onlinecite{Deak11}). Due to the
different choice of the basis, the HSE06 values presented here differ
somewhat from those in Refs.~\onlinecite{Gali09,Deak10,Deak11}, but
tests on the negatively charge nitrogen-vacancy center (NV) have shown that
the higher cutoff would cause only negligible difference in the
equilibrium geometry of that defect too.  Defects in the supercell
were allowed to relax in constant volume till the forces were below
0.01~eV/\AA .

We calculated the hyperfine tensors of $^{13}$C and $^{28}$Si isotopes
within PAW formalism\cite{Blochl00} as implemented in VASP 5.3.2
package. Here, we applied a larger 500(1000)~eV cutoff for the plane
wave (charge density) expansion. The hyperfine tensor ($A^{(J)}_{ij}$) between the
electron spin density $\sigma({\bf r})$ of an electron spin $S$, and the non-zero nuclear spin $I$ of nucleus J may be written as
\begin{equation}
A^{(J)}_{ij}=\frac{1}{2S}\mu_e\mu_J 
\Big[ \frac{8\pi}{3} \int\delta({\bf r}-{\bf R}_J)\sigma({\bf r})d{\bf r}+W_{ij}({\bf R}_J) \Big]
\label{eq:hyperfine}
\end{equation}
where the first term within the square brackets is the so-called (isotropic) Fermi-contact term, and
\begin{equation}
W_{ij}({\bf R})=\int \Big(\frac{3 ({\bf r}-{\bf R})_i ({\bf r}-{\bf R})_j}{|{\bf r}-{\bf R}|^5} - \frac{\delta _{ij}}{|{\bf r}-{\bf R}|^3}\Big) \sigma({\bf r})d{\bf r}
\label{eq:anisotropic}
\end{equation}
represents the (anisotropic) magnetic dipole-dipole contribution to the
hyperfine tensor.  $\mu_J$ is the nuclear Bohr-magneton of nucleus
$J$ and $\mu_e$ the electron Bohr-magneton. The Fermi-contact term
is proportional to the magnitude of the electron spin density at the
center of the nucleus which is equal to one third of the trace of the hyperfine tensor, $\frac{1}{3}\sum_iA_{ii}$.
The Fermi contact term arises from the spin density of unpaired
electrons with $s$-character, and can be quite sizable.  The spin
density built up from unpaired electrons of $p$-character yields the
dipole-dipole hyperfine coupling.  The fraction of Fermi-contact and
dipole-dipole term implicitly provides information about the character
of the wave function of the unpaired electron as well as the
corresponding nuclei (via $\gamma_J$). According to our recent
study\cite{Szasz13} the contribution of the spin polarization of core
electrons to the Fermi-contact hyperfine interaction\cite{Yazyev05} is
significant for $^{13}$C isotopes, thus, we include this term in the
calculation of hyperfine tensors.

The excitation energies were calculated within constrained DFT method
that was successfully applied to NV center in
diamond.\cite{Gali09} In this method one can calculate the relaxation
energy of the nuclei due to optical excitation.

The formation energy of the defect with defect charge state $q$ is defined as
\begin{equation}
E_\text{form}^q = E_\text{tot}^{q} - n_{\text{C}} \mu_\text{C} - \mu_\text{Si} + q E_\text{F} + \Delta V(q) 
\end{equation}
by ignoring the entropy contributions, where $E_\text{tot}^q$ is the total energy of the defect in the supercell, $E_\text{VBM}$ is the calculated energy of VBM in the perfect supercell, $\mu _\text{C}$ and $\mu _\text{Si}$ are the chemical potentials of C and Si atoms in diamond with $n_\text{C}$ number of C atoms in the supercell, $E_\text{F}$ is the chemical potential of the electron, i.e., the Fermi-level, and $\Delta V(q)$ is the correction needed for charged supercells.
The thermal charge transition level between the defect charge states
of $q$ and $q+1$ is the position of the Fermi-level ($E_\text{F}$) in
the fundamental band gap of diamond where the formation energies are
equal in these charge states. 
This condition simplifies to a difference in the total energies
 in their respective charge states as follows,
 \begin{equation}
E_\text{F} = E_\text{tot}^{q}- E_\text{tot}^{q+1} + \Delta V(q) - \Delta V(q+1),
\end{equation}
where $E_\text{F}$ is referenced to the calculated $E_\text{VBM}$. For
comparison of different defect configurations and charge states, the
electrostatic potential alignment and the charge correction scheme of
Lany and Zunger was applied for $\Delta V(q)$.\cite{Lany08} Recently, this scheme was
found to work best for defects with medium localization.\cite{Komsa12}

\section{Results}
\label{sec:results}
We first study the structure and the obtained defect levels by HSE06
calculation. Then, we apply group theory in order to explain the
symmetry of the defect states. We use the calculated thermal
ionization energies, excitation energies and hyperfine couplings to
identify the 1.31-eV and 1.68-eV PL centers. We also discuss the results with comparing them to the experiments.

\subsection{\emph{Ab initio} results and group theory analyses of the
  defect states for neutral SiV defect in diamond}
\label{ssec:hse06}

\begin{figure}
\includegraphics[width=0.95\columnwidth]{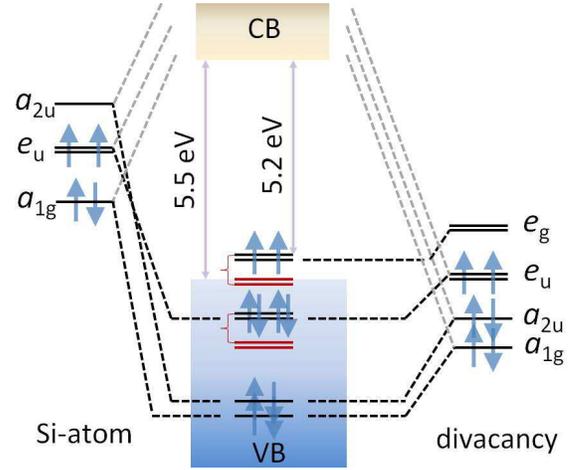}
\caption{\label{fig:leveldiag}(Color online) The defect-molecule
  diagram of the neutral SiV defect in diamond. The irreducible representation of the orbitals under $D_{3d}$ symmetry is shown. The orbitals of the Si-atom and the six carbon dangling bonds of divacancy recombine in diamond where the conduction band (CB) and valence band (VB) are schematically depicted. The $e_g$ orbital is very close to the valence band maximum with forming a strong resonance state at the valence band edge shown as brown (lighter gray) lines. The $e_u$ level falls in the valence band that also forms a strong resonance even deeper in the valence band. For the sake of the simplicity, the position of the spin-up (majority spin channel) levels are depicted in the spin-polarized density functional theory calculation.
}
\end{figure}

\begin{figure*}
\includegraphics[width=0.95\textwidth]{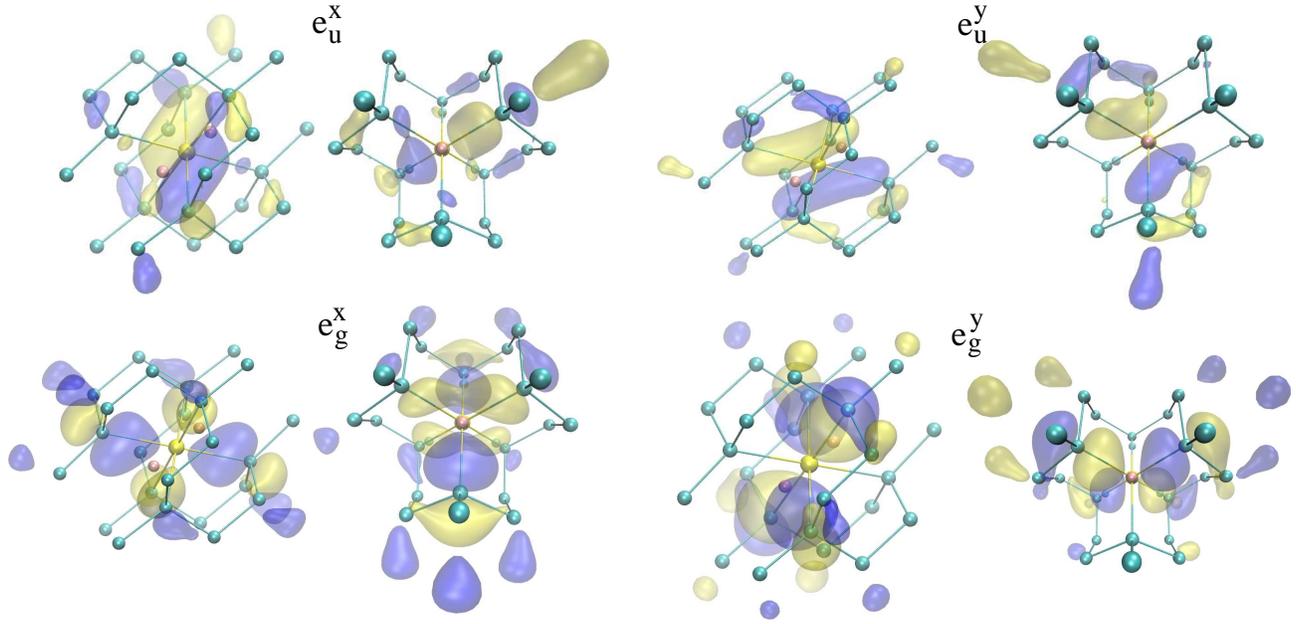}
\caption{\label{fig:states}(Color online) The electron wave function
  of $e_u$ and $e_g$ defect states in the neutral SiV defect. Both
  states are double degenerate, so both $x$ and $y$ components are
  depicted. Note that $e_u$ states are changing sign upon inversion
  while $e_g$ states do not as their irreducible representation
  indicates. Furthermore, Si orbitals have little contribution in
  $e_u$ state but no contribution to $e_g$ state. The $e_u$ state is
  resonant with the valence band, thus less localized, whereas the
  localized $e_g$ state lies in the band gap.  }
\end{figure*}

We first calculated the neutral defect SiV$^0$ by substituting the C-atom
by a Si-atom adjacent to a nearby vacancy. The Si automatically left
the substitutional site creating a split-vacancy configuration which may
be described as a Si-atom placed in a divacancy where the position of
the Si-atom is equidistant from the two vacant sites (see
Fig.~\ref{fig:struct}). The position of Si-atom is a bond center
position which is an inversion center of perfect diamond lattice. In
our special coordinate frame the two vacant sites reside along the
[111] direction of the lattice which has a $C_3$ rotation axis. The
symmetry of the defect may be described as $C_{3v} \times i$ or $D_{3d}$,
where $i$ is the inversion. We note that NV center has
$C_{3v}$ symmetry with no inversion. The defect has $S=1$ high-spin
ground state. This finding agrees with the LDA
calculations.\cite{Goss96,Goss07} After establishing the symmetry of
the defect one can apply group theory analysis for this
defect.\cite{Goss96, Goss07, DHJ11}. One can build the defect states
of this defect as an interaction between the divacancy orbitals and
the Si-impurity states.  The Si impurity has 6 immediate neighbor
C-atoms in divacancy. The calculated distance between Si- and C-atoms
is about 1.97~\AA\ in the neutral charge state, which is
longer than the usual Si-C covalent bond (1.88~\AA ). Since C-atoms are more
electronegative than Si-atom the charge transfer from the Si-atom toward
the C-atoms can be relatively large leaving positively charged Si ion behind. The
divacancy has $D_{3d}$ symmetry with six C dangling bonds. These
dangling bonds form $a_{1g}$, $a_{2u}$, $e_u$ and $e_g$ orbitals while
the Si-related four $sp^3$ states should form $a_{1g}$, $e_u$, and
$a_{2u}$ orbitals in $D_{3d}$ crystal field (the explicit form of these orbitals
as a function of $sp^3$ states can be seen in Ref.~\onlinecite{Hepp2013}). Please, note that the $a_{1g}$,
$a_{2u}$ and $e_u$ states may be combined but $e_g$ orbitals should be
pure C dangling bonds state (see Fig.~\ref{fig:leveldiag}). The
bonding and anti-bonding combinations of these states form the defect
states of SiV defect. 
According to HSE06 calculation the occupation of the defect states may
be described as $a_{1g}^2$$a_{2u}^2$$e_u^4$$e_g^2$ which agrees again
with previous LDA calculations.\cite{Goss96,Goss07} Here, the ten
electrons are coming from the six electrons of C dangling bonds and
the 4 $sp^3$ electrons of the Si impurity. As two electrons occupy the
double degenerate $e_g$ state, the high-spin $S=1$ ground state
naturally forms by following Hund's rules. It is important to
determine the position of the defect levels. Interestingly, HSE06
predicts that only $e_g$ appears in the band gap. In the
spin-polarized calculation the occupied $e_g$ state in the spin-up
channel is at $E_\text{VBM}$+0.3~eV. The $e_u$ state is resonant with
the valence band and can be found just 0.64~eV below VBM. The occupied
$a_{1g}$ state lies very deep in the valence band and may
play no important role in the excitation or ionization processes. 
Our estimations indicate that the $a_{2u}$ defect level is too deep in the
VB to be excited by red excitation. However, higher energetic lasers (in
the green and blue spectrum) can excite this state and other states
within the VB with $a_{2u}$ symmetry.  The
empty anti-bonding orbitals fall in the conduction band, and will not
be considered any more. The most important $e_u$ and $e_g$ states are
depicted in Fig.~\ref{fig:states}.
Despite the $e_u$ state lies in the valence band it is still localized around the defect site. The valence band states are strongly perturbed by the presence of the defect. The VBM is triple degenerate in perfect lattice that splits to an upper $a_{1g}$ level and a lower $e_g$ level in the presence of the defect. Since this $e_{g\text{(VBM)}}$ state has the same symmetry as the low-lying $e_g$ defect state, that $e_{g\text{(VBM)}}$ state becomes a defect resonance state. Similar phenomena occurs for the deep $e_u$ defect state as well. The shallow $e_{g\text{(VBM)}}$ state may play an important role in the excitation/de-excitation process of the defect.

We conclude that the neutral SiV defect (i) has $^3A_{2g}$ ground state,
(ii) can be theoretically ionized as (2+), (1+), as well as (1-) and
(2-) by emptying or filling the double degenerate $e_g$ state in the
gap. In order to establish the relevant charge states, one has to
calculate the adiabatic charge transition levels of SiV defect. 

\subsection{Charge transition levels of SiV defect in diamond}
\label{ssec:occlev}

\begin{figure}
\includegraphics[width=0.95\columnwidth]{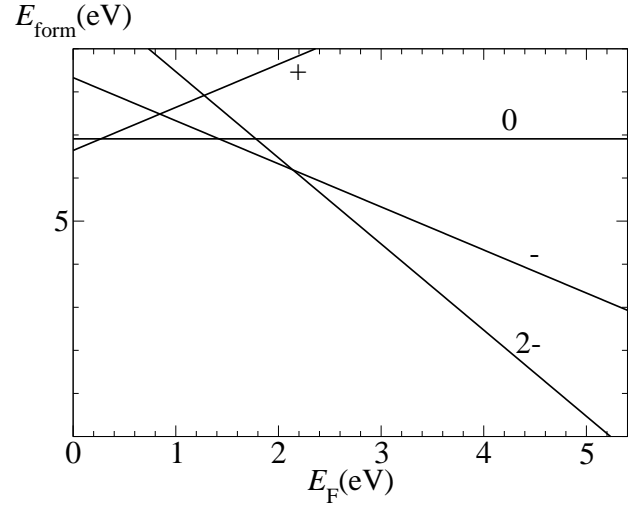}
\caption{\label{fig:occlev} The calculated formation
  energy of SiV defect as a function of the Fermi-level in the
  gap. The crossing points represent the charge transition levels. The
  chemical potential of Si is taken from cubic silicon carbide in the
  carbon-rich limit.
}
\end{figure}

We found that the neutral, negatively charged and double negatively
charged states can be found in diamond (see
Fig.~\ref{fig:occlev}). The single positive charged state might be
only found in very highly p-type doped diamond samples.
The calculated $(-|0)$ level at $E_\text{VBM}$+1.43~eV is very
close to the level that is associated with the acceptor ionization
energy of the defect from photo-conductivity
measurements at $\sim$$E_\text{VBM}$+1.5~eV.\cite{Allers95} Interestingly, the calculated $({2-}|-)$
level at $\sim$$E_\text{VBM}$+2.14~eV is well below the midgap. Since
the $e_g$ level is fully occupied at $E_\text{VBM}$+1.5~eV in $({2-})$ charge state intra-level
optical transition cannot take place, and ultra-violet excitation ($\sim$4.0~eV) is needed to excite or
ionize the defect optically to the conduction band edge.  Another important note that the
calculated acceptor level of NV lies at
$\sim$$E_\text{VBM}$+2.6~eV.\cite{Mizuochi12,Deak13} This means that if both
NV and SiV defects are present in the diamond sample then most of the
NV defects should be neutral in order to detect SiV$^-$
defect. Another important point that the luminescence from SiV$^-$
defect can be more stable than that of NV$^-$ defect in nanodiamonds as
a function of surface termination because the corresponding $({2-}|-)$
charge transition level lies deeper in the band gap than the $(-|0)$
level of NV defect. All in all, the neutral and negatively charged SiV
defects are relevant for intra-level optical transitions. 

\subsection{Negatively charged SiV defect in diamond: analysis of the
  ground and excited states, and the zero-phonon-line energy}

In the negatively charge SiV defect the ground state electron
configuration is $e_u^4$$e_g^3$. This is principally a Jahn-Teller
unstable system as the double degenerate $e_g$ level is partially
filled (in the spin-down channel in our calculation). This has $^2E_g$
symmetry in $D_{3d}$ symmetry. In HSE06 geometry optimization we
allowed the systems to relax to lower symmetries. Indeed, HSE06 showed
a $C_{2h}$ distortion where two C-atoms have 0.03~\AA\ longer distance
from Si-atom than the other two atoms. In this particular case $^2B_g$ state
was formed in $C_{2h}$ symmetry. However, we have to note that the dynamic coupling between
vibrations and electronic states cannot be taken into account in our
calculation. Thus, dynamic Jahn-Teller system cannot be directly
described by our method. For instance, static Jahn-Teller effect
occurs for NV$^0$ in Ref.~\onlinecite{Gali09PRB} while it is known from experiments
that it is a dynamic Jahn-Teller system. Since the distortion from
$D_{3d}$ symmetry obtained by HSE06 calculation is small the defect
may well have $D_{3d}$ symmetry with dynamic Jahn-Teller
effect. We note that no such an electron paramagnetic resonance center with $S=1/2$ spin was found that could be associated with SiV-defect. This also hints that the ground state of SiV$^{-}$ is a dynamic Jahn-Teller system which prohibits the electron spin resonance signal similar to NV$^0$ defect.

\begin{figure}
\includegraphics[width=0.95\columnwidth]{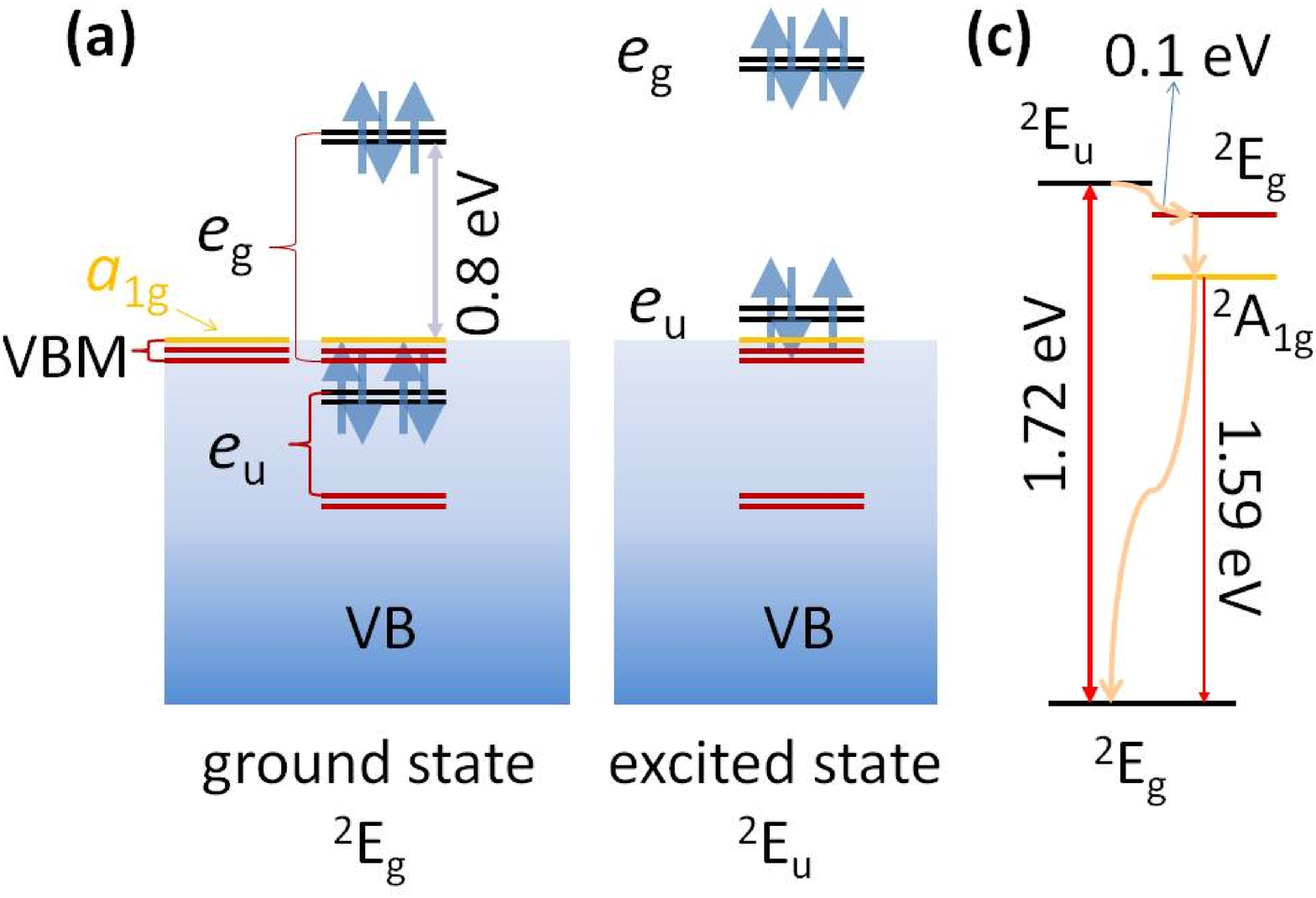}\\
\includegraphics[width=0.95\columnwidth]{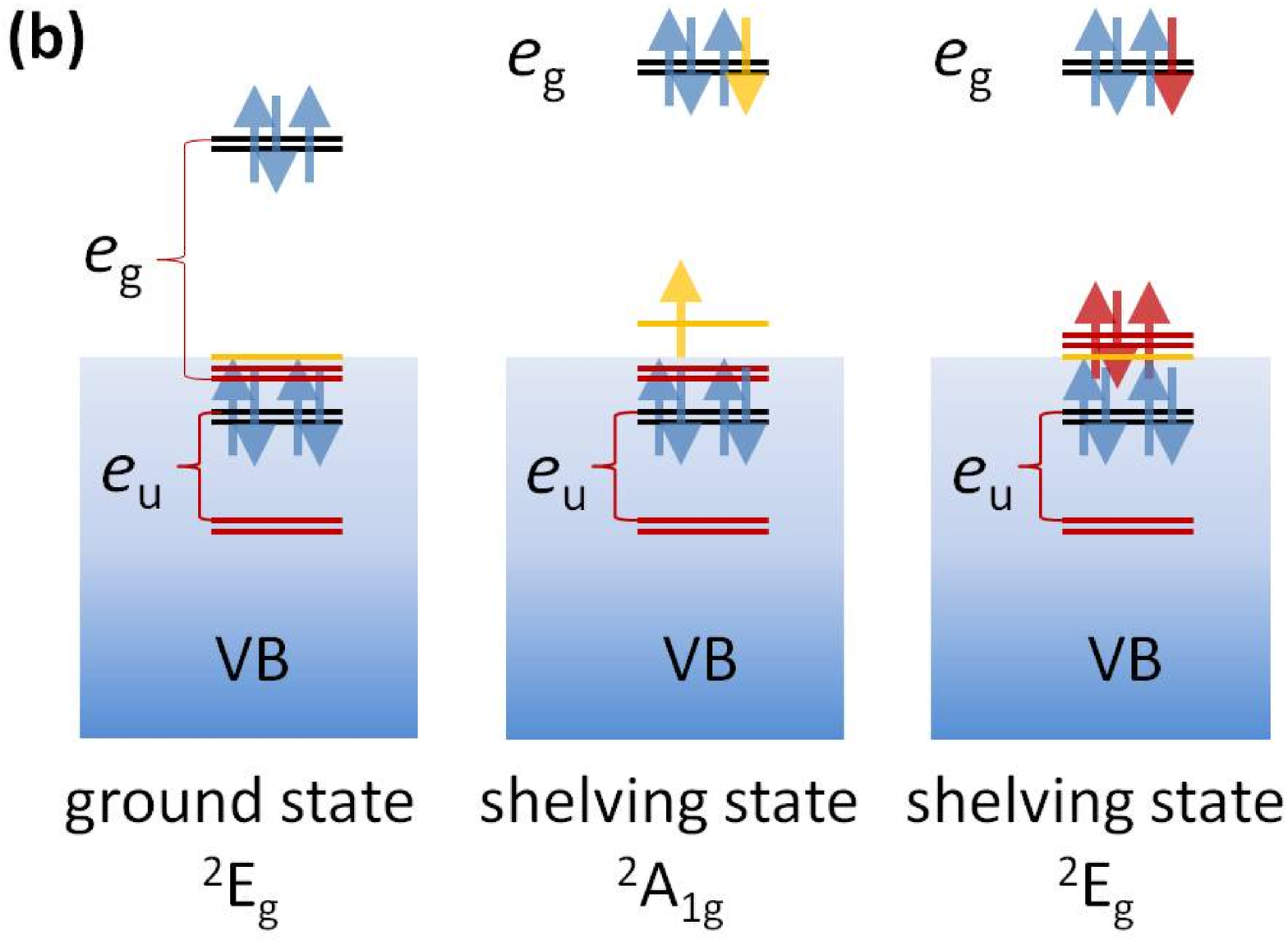}\\
\caption{\label{fig:SiVexc}(Color online) Schematic
  diagram about the electronic structure of the negatively charged SiV defect in diamond. The valence band (VB) resonant states play a crucial role in the excitation.  For the sake of the simplicity, the spin-polarization of the single particle levels are not shown. (a) Ground and the bright (optically allowed) excited state. (b) Ground and shelving states when the hole left on the split valence band edge states. (c) Schematic energy level diagram based on HSE06 constraint density functional theory calculations. Thick red lines indicate strong absorption/emission while the thin red line indicates a weak radiative recombination in the case of the presence of strain which breaks the inversion symmetry of the defect. Orange wavy arrows represent non-radiative recombination down to the ground state. 
}
\end{figure}

Now, we discuss the possible excited states of this system.
Again, the fully occupied $e_u$ state is resonant with the
valence band.  Still, one can promote one electron from this level to
the $e_g$ level in the band gap. The resulted $^2E_u$ excited state is
again Jahn-Teller unstable. Interestingly, when the electron from the minority spin-down $e_u$ level in the valence band was promoted to the $e_g$ state in the gap then the resulted hole state pops up clearly above the valence band edge at about $E_\text{VBM}$+0.12~eV [see Fig.~\ref{fig:SiVexc}(a)]. 
The $^2E_u\rightarrow ^2E_g$ optical
transition is allowed. In HSE06 calculation the excited state has also
$C_{2h}$ distortion but the effect is again small, and can be a
dynamic Jahn-Teller system. The calculated ZPL energy is about 1.72~eV which agrees
well with that of 1.68-eV PL center. The calculated relaxation energy
due to optical excitation is about 0.03~eV which is much smaller
than that of NV center (about 0.21~eV), and explains the small
contribution of the vibration sideband in the emission spectrum.  Thus,
the assignment of 1.68-eV PL center with SiV$^-$ defect is
well-supported by our calculation. 

We argue that the sharp zero-phonon line of SiV$^{-}$ as well as the relative strength of the fine-structure splittings in the ground and excited states can be well-understood by our findings. The $^2E_g$ ground and $^2E_u$ excited states have very similar electron charge densities, where mostly just the phase differs between the two states. As a result, the ions will be subject to similar potentials in their ground and excited states leading to only small change in the geometry due to optical excitation. This is in stark contrast to the case of NV$^{-}$ defect where the electron charge density strongly redistributes upon optical excitation leading to a large Stokes-shift.\cite{Gali09} Another observation is that $^2E_g$ state virtually is not localized at all on Si-atom due to symmetry reasons, however, our projected density of states analysis shows that there is a small contribution from the orbitals of the Si-atom in the $^2E_u$ excited state, allowed by symmetry. This may explain the larger splitting in the fine-structure of the excited state than that in the ground state:\cite{Clark95} the $^2E_u$ state has expected to have larger spin-orbit splitting due to the small contribution of the orbitals of Si-atom than that in $^2E_g$ state where this contribution is missing, as the spin-orbit strength increases with the atomic number to the fourth power.  

Furthermore, our HSE06 calculations reveal that the heavily perturbed valence band edges can play important role in understanding the defect properties as those states lie closer to the $e_g$ state in the band gap than the $e_u$ resonant defect state. Therefore, we calculate the excitation from the split VBM states.  A ${}^2E_g$  and a ${}^2A_{1g}$ excited states are built from the hole left on the $e_{g\text{(VBM)}}$ and $a_{1g\text{(VBM)}}$ states, respectively [c.f., Fig.~\ref{fig:SiVexc}(b)].  We find that the ${}^2E_g$ excited state is $\sim$0.1~eV lower in energy than the bright $^2E_u$ excited state where the calculated Stokes-shift of this transition is again small. The ${}^2A_{1g}$ excited state is $\sim$1.59~eV above the ${}^2E_g$ ground state. 

These results can explain the nature of the shelving state in 1.68-eV PL center. Unlike NV-center the SiV defect has inversion symmetry that plays a crucial role in the de-excitation process. Due to the inversion symmetry  the  ${}^2E_g$ and ${}^2A_{1g}$ excited states are shelving states because they are optically forbidden due to the same parity of their wave function with that of the ground state. Thus, the shelving states have the same spin state as the ground and the bright excited states. Since ${}^2E_g$ shelving state has a level very close to that of the bright ${}^2E_u$ state the non-radiative coupling between these states can be efficient. We further note that a new and weak near-infrared (NIR) transition at $\sim$823~nm (1.52~eV) has been found associated with the negatively charged SiV-defect\cite{Neu12}. The measurements implied that this transition belongs to the same charge as the 1.68-eV transition. This NIR transition was particularly found in ensemble measurements of nanodiamond samples when strain was present in the sample\cite{Neu12}. Our calculations imply [Fig.~\ref{fig:SiVexc}(c)] that this weak radiative transition can be explained by the slightly distorted ${}^2A_{1g}$ excited state. Any distortion of the diamond lattice will break the inversion symmetry of the SiV defect, so the parity of the corresponding wave functions. Therefore, transitions between $^2A_1$ and $^2E$ (originally $^2A_{1g}$ and $^2E_g$) will be allowed. The calculated transition energy (1.59~eV) is very close to the detected one which further supports our assignment. Our calculations highlight the importance of the valence band edge states in understanding the optical properties of 1.68-eV center in diamond.

We note again that the $e_u$ state is resonant with the
valence band. Thus, unlike the case of NV center with well-separated
atomic-like states in the gap, it is probable that ionization of
the defect can occur during optical excitation. In this process, a
hole is created that might (temporarily) leave the defect with
creating an optically inactive ($2-$) charge state, particularly, in the presence of external perturbations that create an effective electric field. This ($2-$) charge state can be optically converted back to ($-$) state only by ultra-violet excitation.

\subsection{Neutral SiV defect: hyperfine tensors and zero-phonon-line
energy}

The ($2-$) charge state is a closed-shell singlet while the neutral and
negatively charged SiV defects have $S=1$ and $S=1/2$ spin states, respectively.
The SiV$^0$ defect was assigned to KUL1 EPR center with $S=1$ state
and $D_{3d}$ symmetry.\cite{Goss07, Edmonds08}. Our calculations
supports this assignment (see Table~\ref{tab:hf}). The spin density is
localized on four C dangling bonds in the $e_g$
orbital of divacancy. While the $e_g$ state is not localized on Si
impurity but the spin density from these C dangling bonds can overlap
with the Si atom which promotes a well measurable Fermi-contact term
on the $^{29}$Si nuclei. The hyperfine interaction with $^{29}$Si is
almost isotropic unlike the case of $^{13}$C isotopes which
show typical anisotropic signal due to $sp^3$ dangling bonds.
\begin{table}
\begin{ruledtabular}
\caption{\label{tab:hf} Calculated and measured hyperfine constants
  ($A$) for KUL1 EPR center and neutral silicon-vacancy defect in
  diamond. The KUL1 EPR center data was taken from
  Ref.~\onlinecite{Edmonds08}. Previous theoretical results are taken
  from Ref.~\onlinecite{Goss07} carried out by LDA 216-atom supercell
  calculation without taken into account the core polarization
  (LDA-nocp). The present theoretical values are obtained by HSE06
  functional in 512-atom supercell where the contribution of the
  spin-polarization of core electrons to the Fermi-contact term in $^{13}$C is 26\% of the
  total (HSE06-cp). $\Theta$ is the angle between the symmetry axis
  ([111] direction) and the parallel component of the hyperfine
  constant $A_{||}$.}
 \begin{tabular}{ lccc}
  Interaction   & $A_{||}$ (MHz) & $A_\perp$ (MHz) & $\Theta$ ($^\circ$)  \\
   \hline
  $^{13}$C (experiment)$^a$ & 66.2 & 30.2 & 35.3 \\
  $^{13}$C (LDA-nocp)$^b$ & 51 & 12 &  33.3 \\
  $^{13}$C (HSE06-cp)    & 68 & 28 & 34.6 \\
$^{29}$Si (experiment)$^a$ & 76.3 & 78.9 & 0 \\
  $^{29}$Si (LDA-nocp)$^b$ & 78 & 82 & 0 \\
  $^{29}$Si (HSE06-cp)    &  92 & 97  & 0  \\
 \end{tabular}
\end{ruledtabular}
 \begin{tablenotes}
  \item $^a$ Ref.~\onlinecite{Edmonds08}, $^b$ Ref.~\onlinecite{Goss07}
 \end{tablenotes}
\end{table}
The calculated and measured $^{13}$C hyperfine tensor agree very
well. We note that the contribution of the spin-polarization of core
electrons to the Fermi-contact term is very significant which
\emph{compensates} the hyperfine interaction due to valence
electrons. We found this effect also for NV and other related
defects.\cite{Szasz13} 

Next, we discuss the optical transitions for SiV$^0$. Intra-level
transition may occur between the fully occupied $e_u$ state and the
empty $e_g$ state in the spin-down channel. While the fully occupied
$e_u$ state lies in the valence band a strong resonant excitation may
occur from this state. In hole picture the excited state may be
described as $e_u^1$$e_g^1$ while the ground state is $e_g^2$.
The electron configuration of $e_g^2$ results in ${}^3A_{2g}$ triplet state and
 ${}^1E_{g}$, ${}^1A_{1g}$ singlet states. The ground state is ${}^3A_{2g}$. The
 electron configurations of $e_u^1$$e_g^1$ results in ${}^3A_{1u}$, ${}^1A_{1u}$,
 ${}^3A_{2u}$, ${}^1A_{2u}$, ${}^3E_{u}$, and ${}^1E_{u}$ multiplets. Among these
 states  ${}^3A_{1u}$ and ${}^3E_{u}$ states are optically allowed from the
 ${}^3A_{2g}$ ground state. Polarization studies of the 1.31-eV center indicates that 
 ${}^3A_{1u} \rightarrow {}^3A_{2g}$ transition occurs in the PL process\cite{DHJ11}. The
 ${}^3A_{1u}$ state can be described as linear combination of Slater-determinants that cannot 
be treated within constraint DFT. However, the $M_S$=1 substate of ${}^3E_{u}$ multiplet can 
be described by a single Slater-determinant. The calculated excitation energy is about 1.63~eV 
which is significantly larger than 1.31~eV. Thus, we may conclude that ${}^3E_{u} \rightarrow 
{}^3A_{2g}$ transition is not responsible for the 1.31-eV PL line. As the only other feasible transition is ${}^3A_{1u} \rightarrow {}^3A_{2g}$, our result indirectly confirms the experimental finding. 

We also study the role of the valence band resonant states similar to the case of the negatively charged defect. When the electron is promoted from $a_{1g\text{(VBM)}}$ state to the $e_g$ state in the gap then ${}^3E_{g}$ state is created. The calculated excitation energy of this shelving state is $\sim$1.27~eV. This coincides with the small peak found in Ref.~\onlinecite{DHJ11}, simultaneously with the peak of 1.31-eV. This transition is optically allowed only when the inversion symmetry of the defect is broken. This might happen due to strain in the diamond sample. Nevertheless, our calculation implies that this ${}^3E_{g}$ state can play an important role in the de-excitation process as a shelving or metastable state. 

We note that the experiments indicate\cite{DHJ11} that the ground state of the neutral SiV defect resides at $\sim$0.2~eV  above the valence band edge. According to HSE06 calculations the occupied single particle $e_g$ level lies at $E_\text{VBM}$+0.3~eV that agrees nicely with the implications from the experiments. In addition, we found a metastable state of the defect which may explain the temperature dependence of the 1.31-eV PL spectrum\cite{DHJ11}. We also propose that strain may induce a weak PL transition from this shelving state because this shelving state has the same spin as that of the ground state, and the parity selection rule may be relaxed in slightly distorted geometry of the SiV defect. We further note that
a very similar conclusion can be drawn for the optical excitation of
the neutral SiV defect as was hinted for the negatively charged defect: a hole is created in the valence band in the excitation process that may leave defect temporarily or permanently which leads to a charge conversion from neutral to negatively charged SiV defect. 

\section{Summary and Conclusion}
\label{sec:summary}

We carried out \emph{ab initio} supercell calculations on SiV defect in diamond. The calculations could positively confirm the assignment of 1.68-eV and 1.31-eV PL centers to the negatively charged and neutral SiV defect, respectively. The calculations reveal the high importance of the inversion symmetry of the center as well as the role of resonant valence band states in understanding the optical properties of the defect. We show that the shelving states of 1.68-eV PL center are from valence band excitations where the lowest-energy shelving state may have NIR emission to the ground state in strain diamond samples. We show that holes are created in the excitation of the negatively charged and neutral SiV defects that may lead to charge conversion of these centers. In addition, we show that the acceptor level of SiV defect lies very deep in the band gap. As a consequence, the 1.68-eV PL center can be photo-stable when it is close to the surface of hydrogenated diamond surface. On the other hand, SiV defect is double negatively charged in such diamond samples where the NV-defect is negatively charged. Thus, the Fermi-level should be set lower than the midgap of diamond, in order to conserve its negative charge state. 

\section*{Acknowledgments} 
Discussions with Christoph Becher are highly appreciated. Support from EU FP7
project DIAMANT (Grant No.\ 270197) is acknowledged. J.R.M. acknowledges support from Conicyt PIA program ACT1108.



%

\end{document}